\documentclass[11pt,twoside]{article}

\usepackage{asp2006}
\usepackage{epsf}
\usepackage{psfig}
\usepackage{lscape}

\begin{document}
\title{Optical light curves of RS Oph (2006) 
and hydrogen burning turnoff}   
\author{Izumi Hachisu$^1$, Mariko Kato$^2$, Seiichiro Kiyota$^3$,
Katsuaki Kubotera$^3$, Hiroyuki Maehara$^3$, Kazuhiro Nakajima$^3$, 
Yuko Ishii$^4$, Mari Kamada$^4$, Sahori Mizoguchi$^4$, Shinji Nishiyama$^4$,
Naoko Sumitomo$^4$, Ken'ichi Tanaka$^4$, Masayuki Yamanaka$^4$, Kozo
Sadakane$^4$}   
\affil{$^1$~Dept. of Earth Sci. \& Astr., Univ. of Tokyo,
Tokyo 153-8902, Japan 
~$^2$~Keio Univ. ~$^3$~Var. Star Obs.
League in Japan ~$^4$~Osaka Kyoiku Univ.}    

\begin{abstract} 
We report a coordinated multi-band photometry of the RS Oph
2006 outburst and highlight the emission line free $y$-band photometry that
shows a mid-plateau phase at $y \sim 10.2$ mag from day 40 to day 75
after the discovery followed by a sharp drop of the final decline.
Such mid-plateau phases are observed in other two recurrent novae,
U Sco and CI Aql, and are interpreted as a bright
disk irradiated by the white dwarf.  We have calculated
theoretical light curves based on the optically thick wind theory
and have reproduced the early decline, mid-plateau phase, and final
decline.  The final decline is identified with the end of steady
hydrogen shell-burning, which turned out at about day 80.
This turnoff date is consistent with the end of a supersoft X-ray phase
observed with {\it Swift}.
Our model suggests a white dwarf mass of $1.35 \pm 0.01 ~M_\odot$,
which indicates that RS Oph is a progenitor of Type Ia supernovae.
We strongly recommend the $y$-filter observation of novae to detect
both the presence of a disk and the hydrogen burning turnoff.
Observational data of $y$ magnitudes are provided here together with
other multi-wavelength light curve data. 
\end{abstract}



\section{Optical observation of RS Oph (2006)}
It is well known that the X-ray turnoff time is a good indicator of the
white dwarf (WD) mass \citep[e.g.,][]{hac06ka, hac06kb}.
When the hydrogen shell-burning atop the WD extinguishes,
a supersoft X-ray phase ends \citep[e.g.,][]{kra96}.  In
a visual light curve, however, this turnoff is not clear because
many strong emission lines contribute to it.  To avoid such
contamination to the continuum flux, we have observed RS Oph with
the Str\"omgren $y$-band filter.  The $y$-filter is an intermediate
bandpass filter designed to cut the strong emission lines in the wide
$V$ bandpass filter, so that its light curve represents the continuum
flux of novae \citep[e.g.,][]{hac06kb}.  We have further modeled
the light curve of RS Oph and have estimated the WD mass by fitting
our modeled light curve with the observation.

Optical observations were started just after the discovery of the 2006
outburst \citep{nar06}.  Each observation
are listed in Tables 1 and 2.  We have put a special
emphasis on  the Str\"omgren $y$ filter to avoid contamination
by the strong emission lines.  These $y$ filters were made
by Custom Scientific, Inc.,\footnote{See http://www.customscientific.com.}
and distributed to each observer by one of the authors (M. Kato).
Kiyota, Kubotera, Maehara, and Nakajima (VSOLJ members) started observation
on 2006 February 13 and obtained data for $y$ magnitudes from 2006
February 17 until 2007 June 20.
The Osaka Kyoiku University (OKU) team obtained $V$ and $y$ magnitudes
of 42 nights starting from 2006 February 17 until 2006 November 17.
The magnitudes of this object were measured by using the local
standard star, TYC2 5094.92.1 (Kiyota) or TYC2 5094.283.1
(the other observers).
We adapted the brightness and color of ($y= V= 9.57$, $B-V= 0.56$) for
TYC2 5094.92.1 and ($y= V= 9.35$, $B-V=1.23$) for TYC2 5094.283.1 from
the Tycho2 catalog.

\begin{figure}[!h]
\plotone{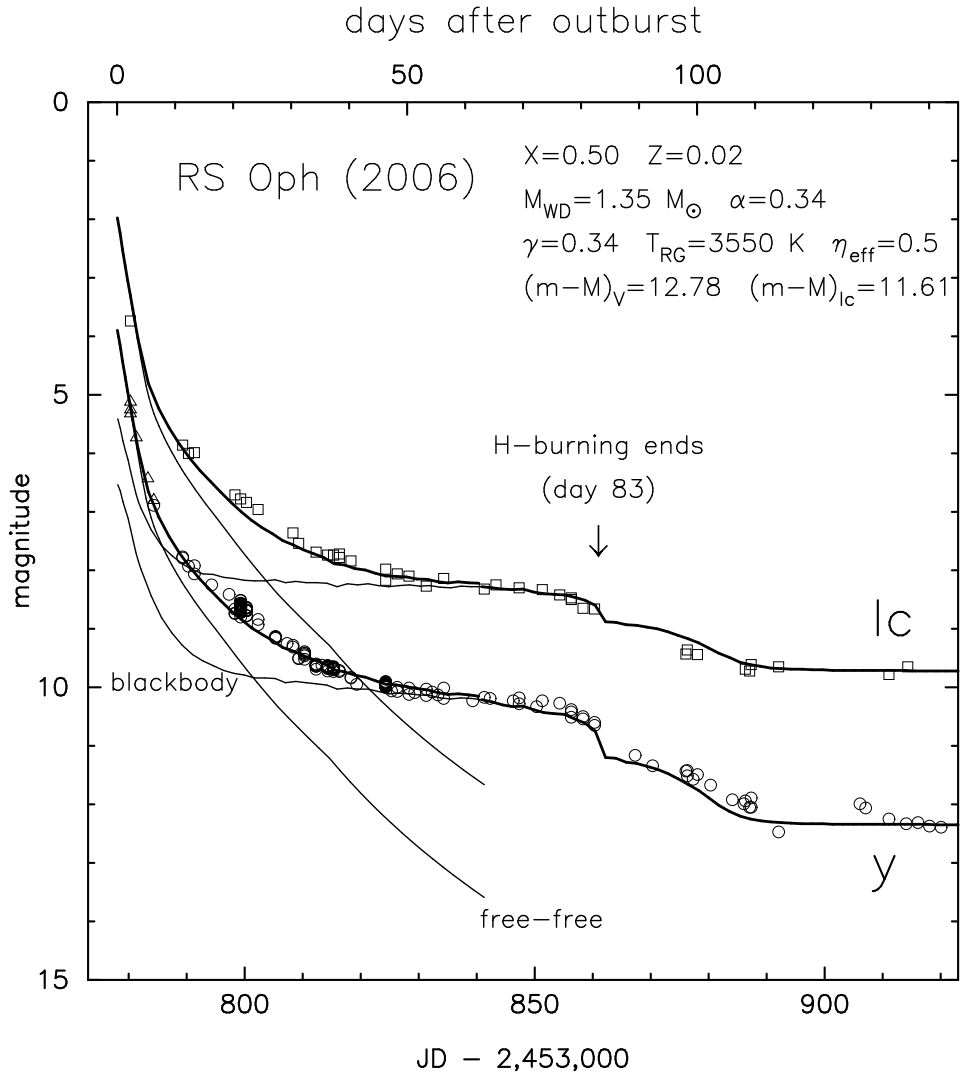}
\caption{ $y$ ({\it open circles})and $Ic$ ({\it open squares})
light curves for the RS Oph (2006).  Thick solid lines are
combined free-free and blackbody.  
}
\end{figure}

\begin{table}[!ht]
\label{oku-data}
\caption{Optical Data of Osaka Kyoiku Univ.}
\smallskip
\begin{center}
{\tiny
\begin{tabular}{llllllll}
\tableline
\noalign{\smallskip}
Days & JD & $y$ & err. & obs. & $V$ & err. & obs.\\
\noalign{\smallskip}
\tableline
\noalign{\smallskip}
 6.33 & 3784.33 & 6.98 & 0.02 & 11 & 6.47 &  0.01 &  6 \\
16.30 & 3794.30 & 8.26 & 0.01 & 14 & 7.61 &  0.01 &  5 \\
24.25 & 3802.25 & 8.83 & 0.01 & 11 & 8.02 &  0.02 & 10 \\
31.30 & 3809.30 & 9.48 & 0.01 & 10 & 8.76 &  0.01 & 10 \\
34.27 & 3812.27 & 9.65 & 0.01 & 10 & 8.90 &  0.02 & 10 \\
37.24 & 3815.24 & 9.69 & 0.02 & 10 & 8.94 &  0.03 & 10 \\
41.25 & 3819.25 & 9.92 & 0.01 & 10 & 9.22 &  0.03 & 10 \\
48.22 & 3826.22 & 10.08 & 0.01 & 10 & 9.39 &  0.02 & 10 \\
51.21 & 3829.21 & 10.07 & 0.01 & 11 & 9.45 &  0.01 & 10 \\
55.22 & 3833.22 & 10.11 & 0.01 & 10 & 9.54 &  0.01 & 10 \\
72.19 & 3850.19 & 10.20 & 0.01 & 10 & 9.80 &  0.02 & 10 \\
73.19 & 3851.19 & 10.28 & 0.02 & 10 & 9.82 &  0.01 & 10 \\
78.14 & 3856.14 & 10.36 & 0.02 & 10 & 9.98 &  0.02 & 10 \\
80.13 & 3858.13 & 10.54 & 0.01 & 10 & 10.08 &  0.01 & 10 \\
83.13 & 3861.13 & 10.68 & 0.03 & 10 & 10.23 &  0.02 & 10 \\
98.09 & 3876.09 & 11.47 & 0.03 & 10 & 11.00 &  0.01 & 10 \\
99.13 & 3877.13 & 11.47 & 0.01 & 10 & 11.03 &  0.01 & 10 \\
102.03 & 3880.03 & 11.52 & 0.02 & 11 & 11.10 &  0.03 & 10 \\
108.08 & 3886.08 & 11.88 & 0.07 & 10 & 11.40 &  0.02 & 10 \\
112.17 & 3890.17 & 11.93 & 0.04 & 10 & 11.45 &  0.02 & 10 \\
115.17 & 3893.17 & 11.85 & 0.04 & 10 & 11.38 &  0.02 & 10 \\
122.15 & 3900.15 & 11.80 & 0.03 & 10 & 11.41 &  0.02 & 10 \\
128.16 & 3906.16 & 11.99 & 0.06 &  5 & 11.56 &  0.11 &  4 \\
152.04 & 3930.04 & 12.01 & 0.06 & 10 & 11.71 &  0.03 &  9 \\
153.15 & 3931.15 & 11.97 & 0.03 & 10 & 11.69 &  0.04 & 10 \\
177.01 & 3955.01 & 11.68 & 0.02 & 10 & 11.47 &  0.03 & 10 \\
184.04 & 3962.04 & 11.83 & 0.02 & 10 & 11.62 &  0.02 & 10 \\
192.96 & 3970.96 & 11.67 & 0.02 & 10 & 11.48 &  0.02 & 10 \\
220.93 & 3998.93 & 11.53 & 0.02 & 10 & 11.38 &  0.01 & 10 \\
227.97 & 4005.97 & 11.70 & 0.04 & 10 & 11.51 &  0.02 & 10 \\
238.95 & 4016.95 & 11.68 & 0.03 & 10 & 11.49 &  0.03 & 10 \\
247.90 & 4025.90 & 11.69 & 0.03 & 10 & 11.60 &  0.03 & 10 \\
248.90 & 4026.90 & 11.76 & 0.04 & 10 & 11.61 &  0.02 & 10 \\
257.93 & 4035.93 & 12.00 & 0.04 & 48 & --  & --  & -- \\
261.90 & 4039.90 & 11.90 & 0.03 & 10 & 11.70 &  0.05 & 10 \\
264.88 & 4042.88 & 11.67 & 0.02 & 10 & -- & -- & -- \\
266.94 & 4044.94 & 11.54 & 0.07 & 7  & -- & -- & -- \\
267.91 & 4045.91 & 11.61 & 0.16 & 6  & -- & -- & -- \\
268.90 & 4046.90 & 11.50 & 0.03 & 10 & -- & -- & -- \\
269.88 & 4047.88 & 11.33 & 0.03 & 10 & -- & -- & -- \\
270.88 & 4048.88 & 11.30 & 0.02 & 10 & -- & -- & -- \\
278.86 & 4056.86 & 11.42 & 0.05 & 10 & -- & -- & -- \\
\noalign{\smallskip}
\tableline\\
\end{tabular}
}
\end{center}
\end{table}

\begin{table}[!ht]
\label{vsolj-data}
\caption{Optical Data of VSOLJ}
\smallskip
\begin{center}
{\tiny
\begin{tabular}{llllllll}
\tableline
\noalign{\smallskip}
Days & JD & $B$ & $V$ & $y$ & vis. & $R$ & $I_c$\\
\noalign{\smallskip}
\tableline
\noalign{\smallskip}
1.36 & 779.36 & - & - & - & 4.46 & - & - \\
2.33 & 780.33 & - & 5.23 & - & 5.11 & 4.46 & 3.74 \\
3.36 & 781.36 & - & 5.73 & - & 5.61 & - & - \\
4.33 & 782.33 & - & - & - & 5.70 & - & - \\
5.36 & 783.36 & - & 6.43 & - & 6.32 & - & - \\
6.35 & 784.35 & 7.53 & 6.80 & 6.89 & 6.70 & 6.06 & - \\
7.33 & 785.33 & - & - & - & 6.93 & - & - \\
9.33 & 787.33 & - & - & - & 7.10 & - & - \\
10.35 & 788.35 & - & - & - & 7.10 & - & - \\
11.32 & 789.32 & 8.18 & 7.46 & 7.78 & 7.16 & 6.04 & 5.86 \\
12.34 & 790.34 & 8.33 & 7.68 & 7.93 & - & - & 6.00 \\
13.31 & 791.31 & 8.54 & 7.66 & 7.99 & 7.43 & 6.05 & 5.99 \\
15.33 & 793.33 & - & - & - & 8.00 & - & - \\
16.33 & 794.33 & 8.85 & 8.00 & 8.25 & - & 6.11 & - \\
19.29 & 797.29 & 8.99 & 8.29 & 8.41 & - & 6.24 & - \\
20.30 & 798.30 & 9.08 & 8.37 & 8.71 & 8.10 & 6.55 & 6.71 \\
21.30 & 799.30 & 8.97 & 8.38 & 8.63 & 8.10 & 6.61 & 6.78 \\
22.29 & 800.29 & 9.22 & 8.39 & 8.69 & - & 6.97 & 6.84 \\
23.26 & 801.26 & - & - & - & 8.30 & - & - \\
24.27 & 802.27 & 9.21 & 8.53 & 8.88 & - & 6.75 & 6.96 \\
27.30 & 805.30 & - & 8.82 & 9.14 & - & - & - \\
29.31 & 807.31 & 9.68 & 9.03 & 9.25 & - & 6.91 & - \\
30.28 & 808.28 & 9.70 & 8.97 & 9.29 & 8.10 & 7.15 & 7.36 \\
31.26 & 809.26 & 9.91 & 9.18 & 9.51 & - & 7.31 & 7.54 \\
32.29 & 810.29 & 9.89 & 9.08 & 9.44 & - & 7.72 & - \\
34.25 & 812.25 & 10.14 & 9.22 & 9.63 & - & 7.48 & 7.69 \\
35.32 & 813.32 & 10.14 & 9.46 & 9.62 & - & 7.33 & - \\
36.24 & 814.24 & 10.13 & 9.26 & 9.65 & 9.60 & 7.58 & 7.74 \\
37.26 & 815.26 & 10.11 & 9.29 & 9.68 & - & 7.53 & 7.74 \\
38.25 & 816.25 & 10.10 & 9.41 & 9.72 & 9.50 & 7.78 & 7.75 \\
40.27 & 818.27 & 10.30 & 9.56 & 9.84 & - & 7.72 & 7.84 \\
41.28 & 819.28 & 10.44 & 9.80 & 9.95 & 9.40 & 7.64 & - \\
42.27 & 820.27 & - & - & - & 9.50 & - & - \\
46.26 & 824.26 & 10.44 & 9.60 & 9.95 & - & 8.00 & 8.07 \\
47.25 & 825.25 & 10.60 & 9.93 & 10.06 & 9.90 & 7.84 & - \\
48.28 & 826.28 & 10.53 & 9.86 & 10.03 & - & 7.99 & 8.06 \\
50.26 & 828.26 & 10.56 & 9.88 & 10.06 & 9.90 & 8.05 & 8.10 \\
51.24 & 829.24 & 10.64 & 9.93 & 10.09 & 10.10 & 7.93 & - \\
53.23 & 831.23 & 10.61 & 9.84 & 10.08 & - & 8.20 & 8.27 \\
54.22 & 832.22 & 10.70 & 9.95 & 10.08 & - & 8.02 & - \\
55.22 & 833.22 & 10.71 & 9.98 & 10.13 & - & 8.04 & - \\
56.27 & 834.27 & 10.69 & 9.95 & 10.10 & - & 8.21 & 8.14 \\
61.21 & 839.21 & 10.87 & 10.06 & 10.23 & - & 8.23 & - \\
63.18 & 841.18 & 10.72 & 9.89 & 10.17 & - & 8.58 & 8.32 \\
64.22 & 842.22 & 10.86 & 10.07 & 10.19 & 10.20 & 8.23 & - \\
65.26 & 843.26 & 10.77 & 10.00 & - & - & 8.49 & 8.25 \\
68.22 & 846.22 & 10.99 & 10.16 & 10.23 & - & 8.38 & - \\
69.25 & 847.25 & 10.92 & 10.10 & 10.23 & - & 8.46 & 8.30 \\
72.18 & 850.18 & 11.07 & 10.21 & 10.33 & - & 8.44 & - \\
73.26 & 851.26 & 10.93 & 10.08 & 10.23 & - & 8.63 & 8.33 \\
76.27 & 854.27 & 11.04 & 10.17 & 10.27 & - & 8.67 & 8.42 \\
78.24 & 856.24 & 11.13 & 10.26 & 10.44 & - & 8.78 & 8.48 \\
80.17 & 858.17 & 10.97 & 10.33 & 10.52 & - & 8.85 & 8.65 \\
82.25 & 860.25 & 11.36 & 10.49 & 10.62 & 10.80 & 8.89 & 8.66 \\
89.16 & 867.16 & 11.81 & 11.01 & 11.16 & - & 9.29 & - \\
92.16 & 870.16 & 11.99 & 11.13 & 11.34 & - & 9.51 & - \\
98.15 & 876.15 & 12.15 & 11.20 & 11.45 & - & 9.88 & 9.39 \\
99.19 & 877.19 & 12.30 & 11.34 & 11.57 & 11.20 & 9.83 & - \\
\noalign{\smallskip}
\tableline\\
\end{tabular}
}
\end{center}
\end{table}

\setcounter{table}{1}
\begin{table}[!ht]
\caption{Optical Data of VSOLJ (continued)}
\smallskip
\begin{center}
{\tiny
\begin{tabular}{llllllll}
\tableline
\noalign{\smallskip}
Days & JD & $B$ & $V$ & $y$ & vis. & $R$ & $I_c$\\
\noalign{\smallskip}
\tableline
\noalign{\smallskip}
100.08 & 878.08 & 12.24 & 11.14 & 11.49 & - & 10.02 & 9.44 \\
102.13 & 880.13 & 12.40 & 11.43 & 11.67 & 11.10 & 9.93 & - \\
106.12 & 884.12 & 12.64 & 11.69 & 11.92 & - & 10.22 & - \\
108.18 & 886.18 & 12.69 & 11.65 & 11.96 & - & 10.33 & 9.69 \\
109.17 & 887.17 & 12.68 & 11.65 & 11.99 & - & 10.35 & 9.66 \\
110.09 & 888.09 & - & - & - & 11.80 & - & - \\
111.14 & 889.14 & - & - & - & 11.40 & - & - \\
114.09 & 892.09 & 12.79 & 11.48 & 12.47 & - & 10.45 & 9.65 \\
115.12 & 893.12 & - & - & - & 11.70 & - & - \\
127.18 & 905.18 & - & - & - & 11.90 & - & - \\
128.07 & 906.07 & 12.83 & 11.79 & 11.99 & - & 10.41 & - \\
129.06 & 907.06 & 12.99 & 11.86 & 12.06 & - & 10.49 & - \\
133.05 & 911.05 & 13.09 & 11.76 & 12.25 & - & 10.67 & 9.78 \\
136.13 & 914.13 & 13.05 & 12.12 & 12.33 & - & 10.71 & 9.65 \\
137.99 & 915.99 & 13.22 & 12.18 & 12.31 & - & 10.78 & - \\
140.04 & 918.04 & 13.22 & 12.17 & 12.37 & - & 10.77 & - \\
141.99 & 919.99 & 13.22 & 12.21 & 12.39 & - & 10.77 & - \\
152.09 & 930.09 & 13.12 & 11.86 & 12.07 & 12.20 & 10.60 & 9.60 \\
153.18 & 931.18 & 13.17 & 11.76 & 11.97 & - & 10.48 & 9.58 \\
155.00 & 933.00 & 13.13 & - & 12.07 & - & 10.57 & - \\
164.00 & 942.00 & 13.24 & 12.14 & 12.31 & - & 10.83 & - \\
165.00 & 943.00 & 13.28 & 12.01 & 12.19 & - & 10.85 & 9.71 \\
166.00 & 944.00 & 13.24 & 12.08 & 12.26 & - & 10.82 & - \\
167.07 & 945.07 & 13.18 & 11.91 & 12.02 & - & 10.71 & 9.64 \\
168.02 & 946.02 & 13.18 & 12.01 & 12.19 & 11.90 & 10.74 & - \\
172.00 & 950.00 & 13.05 & 11.79 & 11.92 & - & 10.57 & - \\
173.05 & 951.05 & 13.02 & 11.70 & 11.80 & - & 10.49 & 9.44 \\
174.00 & 952.00 & 12.97 & 11.69 & 11.90 & - & 10.54 & 9.47 \\
175.04 & 953.04 & 12.91 & 11.59 & 11.69 & - & 10.44 & 9.39 \\
176.10 & 954.10 & 12.98 & 11.53 & 11.59 & - & 10.41 & 9.37 \\
178.99 & 956.99 & 12.95 & 11.66 & 11.76 & - & 10.49 & - \\
180.03 & 958.03 & 12.94 & 11.56 & 11.74 & - & 10.52 & 9.46 \\
182.15 & 960.15 & - & 11.55 & - & - & 10.49 & 9.44 \\
182.99 & 960.99 & 13.00 & 11.77 & 11.85 & - & 10.58 & - \\
184.00 & 962.00 & 12.97 & 11.71 & 11.81 & - & 10.62 & 9.50 \\
188.11 & 966.11 & 13.00 & 11.71 & 11.86 & - & 10.60 & 9.51 \\
190.00 & 968.00 & 13.04 & 11.79 & 11.87 & - & 10.59 & - \\
192.98 & 970.98 & 12.86 & 11.64 & 11.76 & - & 10.49 & - \\
193.99 & 971.99 & 12.82 & 11.56 & 11.70 & - & 10.43 & - \\
199.95 & 977.95 & 12.88 & 11.68 & 11.79 & - & 10.53 & - \\
201.03 & 979.03 & - & 11.43 & - & - & 10.43 & 9.55 \\
202.02 & 980.02 & 12.92 & 11.85 & 11.88 & - & 10.62 & 9.51 \\
203.06 & 981.06 & 12.95 & 11.77 & 11.92 & 11.90 & 10.64 & 9.51 \\
204.96 & 982.96 & 13.08 & 11.97 & 12.02 & - & 10.75 & - \\
205.94 & 983.94 & 13.18 & 12.02 & 12.13 & - & 10.81 & - \\
210.95 & 988.95 & 13.39 & 12.20 & 12.28 & - & 10.89 & 9.74 \\
216.00 & 994.00 & - & 11.51 & - & - & - & 9.41 \\
218.96 & 996.96 & 12.97 & 11.49 & 11.55 & - & 10.35 & 9.36 \\
219.95 & 997.95 & 12.93 & 11.49 & 11.60 & - & 10.36 & - \\
220.93 & 998.93 & 12.97 & 11.47 & 11.61 & - & 10.35 & - \\
223.92 & 1001.92 & 12.91 & 11.44 & 11.46 & - & 10.33 & 9.31 \\
224.92 & 1002.92 & 12.97 & 11.52 & 11.64 & - & 10.42 & - \\
227.93 & 1005.93 & 12.95 & 11.46 & 11.61 & 11.70 & 10.46 & 9.40 \\
230.94 & 1008.94 & - & - & - & 11.80 & - & - \\
236.90 & 1014.90 & 13.03 & 11.66 & 11.81 & - & 10.51 & - \\
237.94 & 1015.94 & 12.96 & 11.54 & 11.62 & 11.80 & 10.44 & 9.40 \\
238.89 & 1016.89 & - & - & - & 11.90 & - & - \\
239.91 & 1017.91 & 12.89 & 11.50 & 11.53 & - & 10.43 & 9.36 \\
244.93 & 1022.93 & 12.91 & 11.56 & 11.55 & - & 10.44 & 9.37 \\
245.91 & 1023.91 & 12.92 & 11.59 & 11.61 & - & 10.48 & 9.41 \\
\noalign{\smallskip}
\tableline\\
\end{tabular}
}
\end{center}
\end{table}

\setcounter{table}{1}
\begin{table}[!ht]
\caption{Optical Data of VSOLJ (continued)}
\smallskip
\begin{center}
{\tiny
\begin{tabular}{llllllll}
\tableline
\noalign{\smallskip}
Days & JD & $B$ & $V$ & $y$ & vis. & $R$ & $I_c$\\
\noalign{\smallskip}
\tableline
\noalign{\smallskip}
251.90 & 1029.90 & - & - & - & 11.90 & - & - \\
254.88 & 1032.88 & - & - & - & 11.90 & - & - \\
255.88 & 1033.88 & - & - & - & 11.80 & - & - \\
259.90 & 1037.90 & - & - & - & 11.90 & - & - \\
265.88 & 1043.88 & 12.84 & 11.62 & 11.60 & - & 10.59 & - \\
268.90 & 1046.90 & 12.70 & 11.45 & 11.42 & - & 10.42 & 9.44 \\
269.87 & 1047.87 & 12.77 & 11.35 & 11.42 & - & 10.31 & - \\
270.87 & 1048.87 & 12.70 & 11.29 & 11.36 & - & 10.24 & - \\
271.85 & 1049.85 & 12.72 & 11.31 & 11.37 & - & 10.27 & - \\
273.89 & 1051.89 & 12.66 & 11.26 & 11.26 & - & 10.21 & 9.25 \\
274.87 & 1052.87 & 12.78 & 11.29 & 11.35 & - & 10.26 & - \\
276.87 & 1054.87 & 12.88 & 11.38 & 11.51 & - & 10.36 & - \\
330.36 & 1108.36 & - & 11.42 & 11.50 & - & 10.49 & 9.37 \\
336.34 & 1114.34 & - & 11.36 & 11.45 & - & 10.49 & 9.40 \\
337.32 & 1115.32 & - & 11.11 & - & - & 10.35 & 9.29 \\
341.33 & 1119.33 & - & 11.21 & 11.29 & - & 10.34 & 9.30 \\
345.33 & 1123.33 & - & 11.44 & - & - & 10.52 & 9.47 \\
347.32 & 1125.32 & - & 11.61 & 11.70 & - & 10.65 & 9.52 \\
355.35 & 1133.35 & 13.15 & 11.96 & 12.08 & - & 11.01 & 9.85 \\
357.32 & 1135.32 & - & 12.02 & 12.11 & - & 11.15 & 9.96 \\
358.32 & 1136.32 & - & 11.79 & 11.81 & - & 11.01 & 9.81 \\
360.33 & 1138.33 & 12.62 & 11.80 & 11.91 & - & 10.96 & 9.83 \\
361.32 & 1139.32 & - & 11.81 & 11.81 & 12.00 & 11.03 & 9.89 \\
365.32 & 1143.32 & - & 11.36 & 11.42 & - & 10.64 & 9.57 \\
366.35 & 1144.35 & - & - & - & 12.00 & - & - \\
368.35 & 1146.35 & 12.74 & 11.54 & 11.56 & 12.00 & 10.58 & - \\
369.31 & 1147.31 & - & 11.32 & 11.38 & - & 10.54 & 9.43 \\
370.32 & 1148.32 & 12.14 & 11.40 & 11.32 & - & 10.57 & 9.45 \\
372.35 & 1150.35 & 12.60 & 11.34 & 11.38 & - & 10.33 & - \\
374.35 & 1152.35 & 12.74 & 11.44 & 11.45 & - & 10.43 & - \\
375.29 & 1153.29 & - & 11.30 & 11.34 & - & 10.44 & 9.36 \\
377.36 & 1155.36 & - & - & - & 11.70 & - & - \\
378.32 & 1156.32 & 12.30 & 11.15 & 11.17 & - & 10.30 & 9.24 \\
380.30 & 1158.30 & - & 11.12 & 11.24 & - & 10.26 & 9.23 \\
381.30 & 1159.30 & 12.53 & 11.26 & - & - & 10.27 & - \\
382.31 & 1160.31 & 12.67 & 11.46 & 11.49 & 11.80 & 10.53 & 9.38 \\
383.32 & 1161.32 & 12.37 & 11.26 & 11.26 & - & 10.29 & 9.21 \\
388.27 & 1166.27 & 12.74 & 11.51 & 11.58 & - & 10.55 & - \\
389.28 & 1167.28 & 12.69 & 11.52 & 11.56 & - & 10.58 & - \\
390.26 & 1168.26 & 12.71 & 11.50 & 11.55 & - & 10.55 & - \\
391.27 & 1169.27 & 12.94 & 11.68 & 11.59 & - & 10.71 & - \\
394.20 & 1172.20 & 12.65 & 11.75 & 11.54 & - & 10.56 & - \\
400.28 & 1178.28 & 12.63 & 11.45 & 11.47 & - & 10.43 & - \\
409.29 & 1187.29 & - & - & - & 11.60 & - & - \\
416.22 & 1194.22 & 12.63 & 11.45 & 11.49 & - & 10.47 & - \\
418.23 & 1196.23 & 12.66 & 11.35 & 11.37 & - & 10.37 & - \\
422.20 & 1200.20 & 12.45 & 11.24 & 11.20 & - & 10.23 & - \\
424.21 & 1202.21 & 12.53 & 11.25 & 11.26 & - & 10.27 & - \\
432.30 & 1210.30 & 12.56 & 11.20 & 11.17 & - & 10.12 & 9.10 \\
442.25 & 1220.25 & 12.26 & 10.95 & 11.05 & - & 10.02 & - \\
450.14 & 1228.14 & 12.33 & 11.16 & 11.15 & - & 10.22 & - \\
454.28 & 1232.28 & 12.33 & 11.14 & 11.17 & - & 10.23 & - \\
460.24 & 1238.24 & 12.34 & 11.22 & 11.13 & 11.30 & 10.23 & - \\
462.25 & 1240.25 & - & 10.95 & - & - & - & - \\
463.26 & 1241.26 & 12.19 & 10.98 & 10.95 & - & 10.06 & - \\
474.08 & 1252.08 & 12.23 & 10.99 & 10.99 & - & 10.09 & - \\
479.12 & 1257.12 & - & - & - & 11.20 & - & - \\
480.12 & 1258.12 & 12.49 & 11.15 & 11.21 & - & 10.21 & - \\
483.15 & 1261.15 & - & - & - & 10.60 & - & - \\
485.19 & 1263.19 & 12.11 & 10.91 & - & - & - & - \\
489.24 & 1267.24 & - & 11.32 & - & - & - & - \\
494.17 & 1272.17 & 12.57 & 11.40 & - & - & 10.43 & - \\
\noalign{\smallskip}
\tableline\\
\end{tabular}
}
\end{center}
\end{table}

\section{WD mass, Mid-plateau phase, and distance}
Our main results are summarized as follows
\citep[see][for more details]{hac06};

\noindent
{\bf 1.} The $y$-band light curve observed can be divided into
three phases:  (a) the early, fast-decline phase until day 40
after the discovery (we assume here $t_{\rm OB} =$ JD 2,453,778.0 as
the origin of time as the outburst),
(b) a mid-plateau phase at $y \sim 10.2$ mag
from day 40 to day 75, and (c) the final decline starting on day 75.
This mid-plateau phase is first identified by our $y$ light curve.
The $y$ and $Ic$ light curves are shown in Figure 1.

\noindent
{\bf 2.} Assuming that the binary consists of a red giant companion,
a WD, and a disk around the WD, we calculate theoretical light curves
and reproduce the observed $y$ and $I_c$ light curves.  Also we assume
that free-free emission from the WD envelope (optically thin plasma)
dominates in the early decline phase while the irradiated disk
(photospheric) component dominates in the mid-plateau phase.
The fitting results are shown in Figure1 \citep[see also
Figures 2 and 3 of][]{hac06}.

\noindent
{\bf 3.} The decline rate of the free-free light curve and the hydrogen
burning turnoff date depend not only on the WD mass but also on
the chemical composition of the envelope \citep{hac06kb}.
First we have fitted the decline with $M_{\rm WD}= 1.35 ~M_\odot$
assuming a chemical composition of $X=0.5$, and then
examined two other cases of hydrogen content, $X=0.70$ and $X=0.35$,
and found that the best fit models are obtained for $M_{\rm WD}= 1.36$
and $1.34 ~M_\odot$, respectively.  So we conclude that the WD mass is
$1.35 \pm 0.01 ~M_\odot$.

\noindent
{\bf 4.} The turnoff date of steady hydrogen shell-burning is
day $\sim 80$, which is consistent with the end of a
supersoft X-ray phase observed with {\it Swift} \citep{hac07}.
The sharp drop of the final decline phase strongly indicates
the epoch when the hydrogen shell-burning ends.

\noindent
{\bf 5.}  From the light curve fitting, we obtain a white dwarf mass of
$1.35 \pm 0.01 ~M_\odot$, which is close to the Chandrasekhar mass
\citep[$1.38 ~M_\odot$, see, e.g.,][]{nom82}.  This
suggests that RS Oph is a progenitor system of Type Ia supernovae.

\noindent
{\bf 6.} The distance is estimated to be 1.3-1.7 kpc from the $y$ and $I_c$
light curve fittings in the late phase of the outburst.

\noindent
{\bf 7.} Finally, we strongly recommend the $y$-filter observation
of novae to detect both the presence of a disk and the hydrogen
burning turn-off.



\end{document}